\documentclass[prl, twocolumn, preprintnumbers,amsmath,amssymb,superscriptaddress]{revtex4-1}
\usepackage{graphicx}
\usepackage{dcolumn}
\usepackage{bm}
\usepackage{subfigure}
\usepackage{multirow}
\usepackage{amsmath}
\usepackage{color}
\usepackage[font=footnotesize, justification=raggedright,singlelinecheck=false]{caption}
\usepackage{mathrsfs}
\usepackage{mathtools}
\usepackage{relsize}

\begin{document}

\title{ Self-Assembly  of Isostatic Self-Dual Colloidal Crystals }

\author{Qun-Li Lei}
\thanks{These authors contributed equally.}
\affiliation{National Laboratory of Solid State Microstructures and School of Physics, Collaborative Innovation Center of Advanced Microstructures,  Nanjing University, Nanjing, 210093, China}
\affiliation{School of Chemical and Biomedical Engineering, Nanyang Technological University, 62 Nanyang Drive, 637459, Singapore}

\author{Wei Zheng}
\thanks{These authors contributed equally.}
\affiliation{National Laboratory of Solid State Microstructures and School of Physics, Collaborative Innovation Center of Advanced Microstructures,  Nanjing University, Nanjing, 210093, China}
\affiliation{School of Chemical and Biomedical Engineering, Nanyang Technological University, 62 Nanyang Drive, 637459, Singapore}

\author{Feng Tang}
\affiliation{National Laboratory of Solid State Microstructures and School of Physics, Collaborative Innovation Center of Advanced Microstructures,  Nanjing University, Nanjing, 210093, China}

\author{Xiangang Wan}
\affiliation{National Laboratory of Solid State Microstructures and School of Physics, Collaborative Innovation Center of Advanced Microstructures,  Nanjing University, Nanjing, 210093, China}

\author{Ran Ni}
\email{r.ni@ntu.edu.sg}
\affiliation{School of Chemical and Biomedical Engineering, Nanyang Technological University, 62 Nanyang Drive, 637459, Singapore}

\author{Yuqiang Ma}
\email{myqiang@nju.edu.cn}
\affiliation{National Laboratory of Solid State Microstructures and School of Physics, Collaborative Innovation Center of Advanced Microstructures,  Nanjing University, Nanjing, 210093, China}

\begin{abstract}
Self-dual structures whose dual counterparts are themselves possess unique hidden symmetry, beyond the description of classical spatial symmetry groups. Here we propose a strategy based on { a nematic monolayer of} attractive half-cylindrical colloids to self-assemble these exotic structures. { This system can be seen as  a 2D system of semi-disks.} By using Monte Carlo simulations, we discover two isostatic self-dual crystals,  i.e., an unreported crystal with pmg {space-group} symmetry and the twisted Kagome crystal. For the pmg crystal approaching the critical point, we find the double degeneracy of the {full} phononic spectrum at the self-dual point, and the merging of two tilted Weyl nodes into one \emph{critically-tilted} Dirac node. The latter is `accidentally' located on the high-symmetry line. The formation of this unconventional Dirac node is due to the emergence of the critical flat bands at the self-dual point, which are  linear combinations of \emph{finite-frequency} floppy modes. These modes can be understood as  mechanically-coupled self-dual rhomb chains vibrating in some unique uncoupled ways. Our work paves the way for designing and fabricating self-dual materials  with exotic mechanical or phononic properties.
\end{abstract}

\maketitle

\paragraph{Introduction}
Duality describes the hidden relationship between two seemingly different structures or descriptions~\cite{savit1980duality,quevedo1998duality}, e.g., the  graph duality in geometry~\cite{berge2001theory}, Kramers-Wannier duality~\cite{kramers1941},  the electromagnetic duality~\cite{f2013elect} etc. Duality can generate a self-dual point, at which the dual counterpart of the structure is itself, and unique hidden symmetry emerges giving rise to some unusual degeneracy beyond the description of  standard space group theories~\cite{louvet2015,hou2018hidden,hou2013hidden}. Recently, a self-dual point was discovered in an isostatic mechanical network whose average coordination number  is twice of the system dimension, which can be used to realize the  `mechanical  spintronics'~\cite{vitelli2020}.  According to the Maxwell rule~\cite{maxwell1864}, this kind of structures  is at the edge of mechanical stability (isostatic point)~\cite{lubensky2009,lubensky2012,
lubensky2015rev,mao2018rev}. This property has  been utilized to design flexible metamaterials~\cite{bilal2017in,xu2019real,b2017flexible,rev2019t,mcinerney2020hidden} and topological mechanical insulators~\cite{kane2014top,lubensky2016m,
po2016phonon,vitelli2016geared,irvine2018t,mao2019prx}.  { Nevertheless, so far, only twisted Kagome lattice was found to be self-dual~\cite{vitelli2020} and the origin of this duality remains unknown. Searching for new self-dual isostatic structures of different symmetries not only helps uncover the principle of duality in related classical systems, but also paves the way to fabricate exotic mechanical/phononic materials, for which one needs to find new strategies to construct microscopic structures with discrete dual transformations~\cite{vitelli2020,gonella2020symmetry}.  }

In this work, we  show that { a nematic monolayer of}  attractive half-cylindrical colloidal rods can self-assemble into two  isostatic self-dual crystals with discrete freedom degrees, i.e., the twisted Kagome crystal and an unreported crystal with pmg { space-group} symmetry. The hidden duality in the new isostatic structure  induces the double degeneracy of the { full} phononic spectrum at the self-dual point and generate unconventional \emph{critically-tilted} Dirac cones which  are `accidentally' located on the high-symmetry lines in the  Brillouin zone (BZ).  As the system departs from the self-dual point,  each Dirac node divides into two Weyl nodes. We find that the emergence of the critical flat bands at the self-dual point is the key to the formation of the {critically-tilted} Dirac cones. These flat bands are  linear combinations of the  \emph{finite-frequency} floppy modes, which can be understood  as  mechanically-coupled self-dual rhomb chains vibrating in some unique uncoupled ways. Our finding  can not only help build  new self-dual materials with exotic mechanical or phononic properties, but  also provide opportunities  to observe  intriguing phenomena in extreme conditions with simple classical systems.


\begin{figure*}[htbp] 
\resizebox{160mm}{!}{\includegraphics[trim=0.0in 0.0in 0.0in 0.0in]{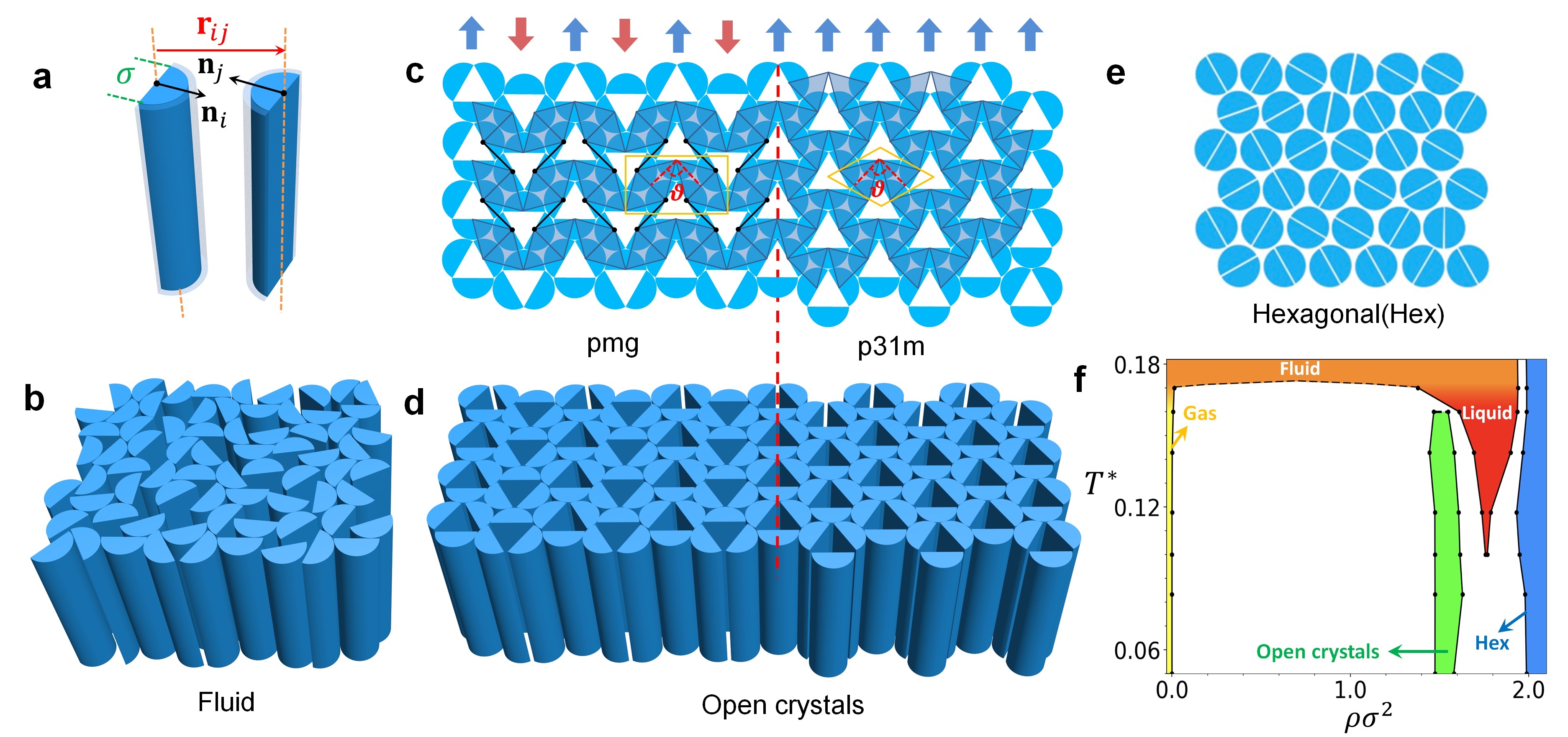} }
\caption{  (\textbf{a}) Schematic of  two parallel half-cylindrical rods. (\textbf{b}) Fluid phase. (\textbf{c,~d}) Top view and side view of open crystals in contact.  Left: pmg crystal. Right: p31m (twisted Kagome) crystal, where $\vartheta$ is the open angle. The blue and red arrows indicate the orientation of the stripes. The orange lines enclose the unit cells of two crystals. (\textbf{e}) Hexagonal (Hex) crystal phase. (\textbf{f})  Phase diagram of the system in dimensions of density $\rho$ and reduced temperature $T^*$.  }
\end{figure*}

\paragraph{Model and Simulation}
We consider { a monolayer of} $N$ hard  half-cylindrical colloidal rods with diameter $\sigma$ (Fig.~1a) and assume the rods are perfectly aligned under external fields or nematic interactions~\cite{sharma2014h, lei2018self}.  Except for the hard core interaction, there is also a short-range attraction between the curved lateral surfaces of two parallel rods, which we model as a square-well potential, with $w=0.1 \sigma$ the square-well width and $\varepsilon$ the well depth. This attraction can be realized by  depletion force~\cite{kraft2012surface} and capillary bridging force~\cite{b2016capillary}  with surface modifications, {or effective interaction arising from nematic liquid crystal elasticity~\cite{m2016triclinic,s2018liquid}}.  We define a reduced temperature $T^{*} = k_{B}T/\varepsilon$, where $k_{B}$ and $T$ are the Boltzmann constant and the temperature of system, respectively. Since the rods are perfectly aligned, the system can be effectively modelled as a { 2D semi-disk system}. $\rho = N/A$ is the 2D density with $A$ the 2D area. We perform Monte Carlo simulations of the system with full periodic boundary conditions~\cite{zheng2020,lei2018self,lei2017}. The details of the simulation can be found in~\cite{Supplemental2}. 

\paragraph{Entropy-stabilized Isostatic Open Crystals}
In Fig.~1f, we show the obtained phase diagram of the system. At high  $T^*$, the fluid phase (Fig.~1b) is a thermodynamically stable phase up to $\rho\sigma^2 \simeq 1.8$, while the hexagonal crystal phase (Fig.~1e) can only be stable when $\rho\sigma^2 \gtrsim 2.0 $. The hexagonal crystal is a hierarchical structure consisting of dimers of two close-packed half-cylinders. With decreasing $T^*$, the fluid phase is divided into a dilute gas phase and a dense liquid phase. The critical regime of the gas-liquid phase separation is indicated by the dashed line in the phase diagram. With further decreasing $T^*$, the open crystal phase (Fig.~1d) appears between the gas and liquid phases. The latter finally disappears when $T^* \lesssim 0.1$.The phase diagram remains qualitatively the same when decreasing the attraction range (see Fig.~S1 in~\cite{Supplemental2}). As shown in Fig.~1c,d, the open crystals  are also hierarchical structures, which are composed of the half-cylindrical trimers with the triangle empty void. These trimers align in stripes whose direction can be up (A) and down (B) as indicated by the blue and red arrows in Fig.~1c. We find the AAAA sequence leads to the twisted Kagome crystal with p31m symmetry, while the ABAB sequence produces an unreported crystal with pmg symmetry.  The unit cells for the two open crystals are drawn in Fig.~1c using orange lines. Similar to the twisted Kagome lattice  which can be obtained from the standard Kagome lattice through twisting the unit cell (Guest-Hutchinson mode~\cite{guest2003determinacy}), the pmg lattice can be formed from a rectangular lattice by doing a similar deformation (see Fig.~S2 for the deformable pmg lattice assembled from Lego units in~\cite{Supplemental2}). More importantly, the densities of both two crystals are controlled by the same open angle $\vartheta$ as shown in Fig.~1c. Therefore, two crystals are compatible with each other on the phase boundary (see the dashed line in Fig.~1c and Fig.~S3 in~\cite{Supplemental2}). The free energy of pmg crystals is only slightly lower than the p31m crystal (about $0.03~k_BT$ per particle) due to the different vibrational entropy, according to free energy calculations based on Einstein integration and dynamic matrix theory (see text and Table S1 in ~\cite{Supplemental2}). This explains why in our direct Monte Carlo simulation, only the  nucleation of random stripe sequences instead of perfect p31m or pmg crystals is observed, similar to the nucleation of FCC/HCP crystal from hard-sphere fluids~\cite{frenkel1984new}. In experiments, one can use a prefabricated template to induce the growth of the preferred crystal~\cite{Hynninen2007,avb1997}.

For open crystals, each particle forms 4 attractive bonds with its neighbours, making them  isostatic crystal according to the Maxwell rule. Therefore, both two open crystals can be stretched to about $150\%$ without energy cost or symmetry breaking  when $T^*\rightarrow 0$ (see Fig.~S4 in~\cite{Supplemental2}). Nevertheless, our simulation shows that the equilibrium open angle is $\vartheta_{\rm eq}=95.7^{\circ}$ at zero pressure ($T^*= 10^{-5}$, $w =10^{-2}\sigma$).  In ~\cite{Supplemental2}, we construct a mean-field theory to calculate the rotational entropy of rods as function of open angle $\vartheta$. This mean-field theory predicts that the rotational entropy is maximized around $\vartheta_{\rm eq}=92.2^{\circ}$ for both two open crystals, close to the measured value. Therefore, these two isostatic crystals are mechanically stabilized by entropy~\cite{mao2015m,hu2018entropy, mao2013nature}. It should be mentioned that $\vartheta_{\rm eq}$ can be effectively tuned in the range from $89.0^{\circ}$ to $95.7^{\circ}$ by changing the attractive area on the half-cylinders (see Fig.~S5 in~\cite{Supplemental2}). As shown later, this range covers the critical open angle $\vartheta_c=90^{\circ}$ at which the structure is self-dual.

\begin{figure}[htbp] 
\centering
\begin{tabular}{c}
	\resizebox{85mm}{!}{\includegraphics[trim=0.0in 0.0in 0.0in 0.0in]{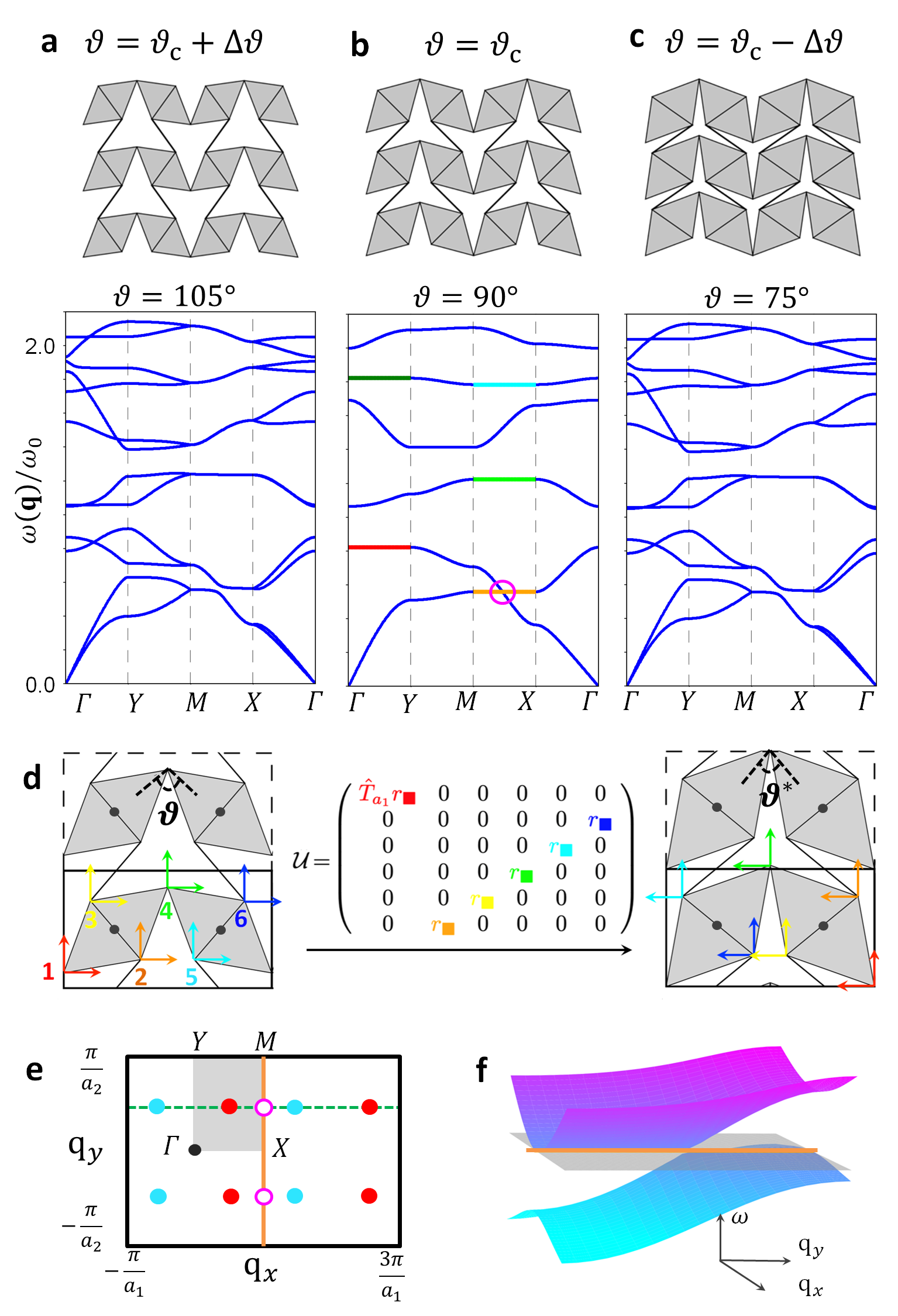} }
\end{tabular}
\caption{  (\textbf{a-c}) Phononic spectrum of pmg hamonic crystals at different open angles. The flat bands are indicated by different colors except blue. the Dirac node is marked by the meganta circle. (\textbf{d}) Duality transformation that maps the vibrational  freedom degrees  between two pmg lattices.  { Black dots are the centers of space inversion.} (\textbf{e}) Two adjacent Brillouin zones separated by the orange line at $q_x=\pi/a_1$. The red (blue) sold points represent the tilted Weyl nodes with Berry phase $\pm \pi$, which merge into a critically-tilted Dirac node (meganta circle) as $\vartheta  \rightarrow  \vartheta_c$. (\textbf{f}) 3D band surfaces of bands (1, 2) and bands (3, 4) near the critically-tilted Dirac node. } \label{band_structure}
\end{figure}

\paragraph{Duality of pmg Crystals}
At  low temperature when particles only vibrate around the  lattice, the open crystal can be described by an effective Hamiltonian under the harmonic approximation (see~\cite{Supplemental2} for the derivation),
\begin{align}
H_{\rm{eff}}&=\sum_{i}\frac{k}{2} {\mathbf u}_{i}^{2}+\frac{\kappa}{2}\sum_{h} \Delta \alpha_{h}^{2} \label{harmonic_approximation}
\end{align}
where  ${\mathbf u}_{i}$  ($ \Delta \alpha_{h}$) is the  translational (rotational)  displacement of particle $i$ (bond angle $h$) from its equilibrium value. Here the first term accounts for central-force attractive bonds and the second term describes the effective bond-bending rigidity arising from particle rotational entropy, with $k$ and $\kappa$ the corresponding effective spring constants. When $T^*\rightarrow 0$ and $w \rightarrow  0$, which corresponds to the low temperature limit for system with general interaction, the bond-bending rigidity can be neglected compared with the central-force bond stiffness. We calculate the phononic spectrum $\omega({\mathbf q} )$ for the pmg crystal under this condition by solving the eigenvalue of the dynamic matrix (see~\cite{Supplemental2} for details). In Fig.~2a,b,c, we show $\omega({\mathbf q} )$ under three different open angles, i.e., $\vartheta_c =90^{\circ}$ and $ \vartheta_c\pm \Delta \vartheta$ with $\Delta \vartheta =15^{\circ}$. We observe the identical phononic spectrum for $\vartheta_c\pm \Delta \vartheta$ and the two-fold degeneracy of the spectrum at the critical angle $\vartheta_c$, similar to what was found in twisted Kagome lattice~\cite{vitelli2020}. This indicates that this critical angle is associated with a hidden symmetry which produces the same effect in spin 1/2 electronic systems according to Kramers’ theorem~\cite{klein1952degeneracy,vitelli2020}. We further prove that there exists a dual transformation represented by the unitary matrix $\mathcal{U}$ between pmg lattices at two different open angles $\vartheta$ and $\vartheta^*$  (Fig.~2d), which satisfies
 \begin{align}
\mathcal{U}({\mathbf q}) D(\vartheta^*, -{\mathbf q}) \mathcal{U}^{-1}({\mathbf q}) = D(\vartheta, {\mathbf q})  \label{dual_trans}
 \end{align}
with $\vartheta^* = 2\vartheta_c - \vartheta$. { Therefore,  $\vartheta=\vartheta^*=\vartheta_c$ is the self-dual point.} Here $D(\vartheta, {\mathbf q}) $ is the dynamic matrix of the system at open angle $\vartheta$.  In $\mathcal{U}$, different rows with different colors correspond to different sites in the unit cell. One can see that $\mathcal{U}$ contains site exchange operation ($2\rightarrow 6$, $3\rightarrow 5$). It also contains the translation operator  ${\hat T}_{x} =  e^{-i {\mathbf q}\cdot {\mathbf a}_1 } $ that shifts the site 1 by one period in  $x$ direction. Lastly, $\mathcal{U}$ contains four-fold rotation operators $r_{ \mathsmaller \blacksquare}  =
\begin{pmatrix}
0 & 1\\ -1  & 0
\end{pmatrix} $ acting on the vibrational freedom degrees. 

\paragraph{Critically-Tilted Dirac Cone}
In Fig.~2a-c, the eigenmodes are double degenerated along the $MX$ line in BZ (Fig.~2e) at arbitrary $\vartheta$. This degeneracy is due to the non-symmorphic glide reflection symmetry in $x$ direction. With $\vartheta \rightarrow \vartheta_c$, the double-degenerated bands (1, 2) and bands (3, 4) along the $MX$ line begin to contact at a point (the magenta circle) and form a completely flat band (orange line). The linear behaviour around this point suggests that it is a Dirac node composed of two coincident Weyl nodes. To confirm this, we calculate the  Berry phase
\begin{eqnarray}
\gamma_j = i \oint_{\mathcal{C}} d {\mathbf q} \cdot \overrightarrow{\mathbf u}_j({\mathbf q}) \nabla_{{\mathbf q}} \overrightarrow{\mathbf u}_j^{\dagger} ({\mathbf q})
\end{eqnarray}
along an enclosed trajectory $\mathcal{C}$ round the Dirac node in the BZ for band $j$, where $\overrightarrow{\mathbf u}_j ({\mathbf q})$ is the eigenstate at $\mathbf q$. For 2D Dirac (Weyl) systems, the Berry phase divided by $\pi$ is a quantized topological number (winding number)~\cite{de2012man,g2017dirac}.  The Weyl nodes with Berry phase $\pm \pi$ can be viewed as topological charges or vortices with sign $\pm$ in the vector field of 2D momentum space~\cite{de2012man}. We find that the Berry phase is $\pm \pi$ for either bands (1,~2) or bands (3,~4) when the trajectory encloses the node, while the summation of the two phases remains zero, indicating that topological charges neutralize at the Dirac node.  On the contrary, the Berry phase remains zero  when the trajectory does not enclose the nodes for all four bands.  In Fig.~2e, we also plot the 3D band surfaces of bands (1, 2) and bands (3, 4). We find that the contact point essentially connects two double degenerated critically-tilted Dirac cones. The flat band along the $MX$ line turns out to be the tangent line (orange line) of two bands with a horizontal surface (shadow plane). We notice that similar critically-tilted Dirac (Weyl) cones,  which are also called  Type III  or zero-index Dirac (Weyl) cone, were recently discovered in electronic and phononic systems~\cite{Bergholtz1,Bergholtz2,liu2018p,huang2018black, m2019type,z2020zero}. At this cone, the quasiparticles have highly-isotropic mass, which can induce many interesting phenomena, like zero flection index~\cite{z2020zero}, black hole horizon~\cite{huang2018black,liu2020fermionic} and enhanced superconducting gap~\cite{li2017effect} etc. In our system the Dirac nodes is protected by glide reflection and the hidden self-dual symmetry, thus  changing the mass of particle pair (1,~4), (2,~6) or (3,~5) does not lift the Dirac nodes.  Morever, the Dirac nodes are  located  `accidentally' on a high-symmetry lines. Ref.~\cite{hou2018hidden,hou2013hidden} showed that the accidental degeneracy can result from the hidden antiunitary symmetry, which is consistent with the antiunitary nature of dual transformation~\cite{vitelli2020}. 

As the system departing from the critical point, each Dirac node is separated into two non-critically-tilted Weyl nodes with opposite Berry phases $\pm \pi$, which we represent as red (+) and blue (-) solid points in Fig.~2e. These nodes are the contact points between bands (2, 3) which are hidden in Fig.~2a,c. The dividing of one Dirac node into two Weyl points in our system is similar to the  transition from Dirac  semimetals to Weyl semimetals in electronic systems~\cite{wan2011t}. In 3D, the Weyl nodes can be `accidentally' generated at  low-symmetry points in BZ~\cite{wan2011t}, while in 2D it would be impossible without additional symmetry constraints. {In our system, we find that the Weyl nodes can be lifted by changing the mass of the particle pair (2,~6), (3,~5), (2,~5) or arbitrary single particle, but not for the particle pair (1,~4), (2,~3) or (5,~6).  This suggests these nodes are protected by the space inversion symmetry with the center points marked as black dots in Fig.~2d. }

\begin{figure}[htbp] 
\centering
\begin{tabular}{c}
	\resizebox{90 mm}{!}{\includegraphics[trim=0.0in 0.0in 0.0in 0.0in]{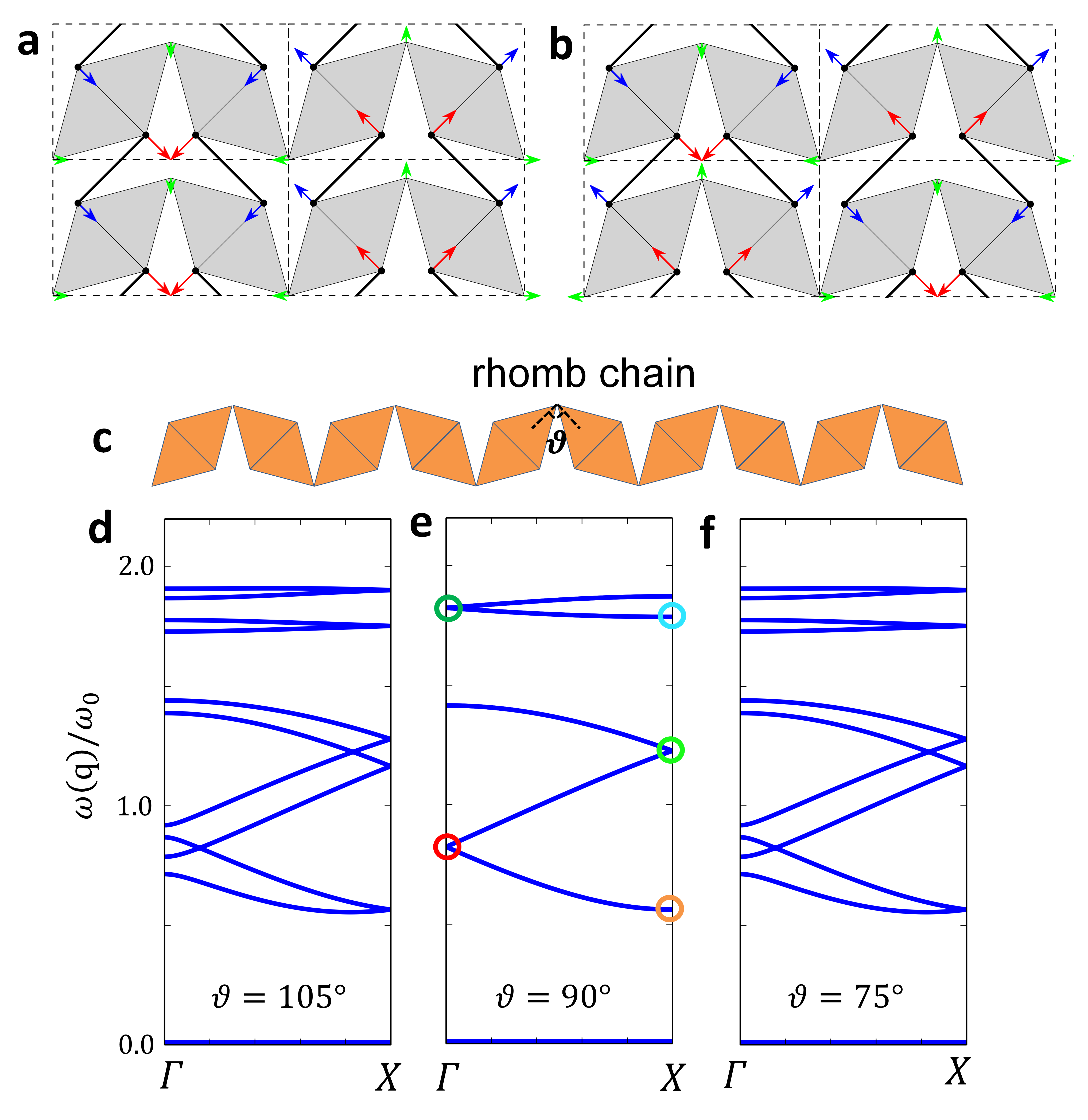} }
\end{tabular}
\caption{  (\textbf{a,b}) Vibrational modes  at  $X$  and $M$  points in BZ for the orange flat band along the $XM$ line.  The vibration direction of linking points (red and blue arrow) are perpendicular to the linking bonds (thick black lines).  (\textbf{c}) The structure of the chain of rhombic units. (\textbf{d-f}) Phononic spectrum of the chain  at different open angles. The color circles at $\Gamma$ and $X$ point correspond to the  flat bands of the same colors in Fig.~2b. }
\end{figure}

\paragraph{Critical Flat Bands and Finite-Frequency Floppy Modes}
When $\vartheta$ approaches $\vartheta_c$, another interesting phenomenon is the emergence of double-degenerated flat bands which are indicated by five different colors except blue in Fig.~2b. As shown previously, the orange flat band is the key to the formation of the critically-tilted Dirac cone. In Fig.~3a,b, we show  one of two eigenmodes for $M$ point and $X$ point on this flat band respectively. Their degenerated counterparts are shown in Fig.~S6 in~\cite{Supplemental2}. In the~\cite{Supplemental2}, we prove that these flat bands are linear combinations of finite-frequency floppy modes in which vibration freedom degrees of linking bonds (thick lines in Fig. 3a,b) are frozen. Since these linking bonds contribute zero energy, these floppy modes are also the eigenmodes of the chain of rhombic units (see Fig.~3c).  In Fig.~3d,e,f, we plot the phononic spectrum of this chain system at three different open angles $\vartheta$, as the same as that in Fig.~2a,b,c.  We find the duality of the 2D lattice is preserved in this quasi-1D system,  which results in the same double degeneracy of band structure at $\vartheta_c$. Exact mapping can be built between the  flat bands in Fig.~2b and the degenerated points in Fig.~3e, which are marked by circles of the same colors.   These findings suggest the possibility of designing high-dimensional  self-dual structures from the low-dimensional ones~\cite{chen2016topological}.

\paragraph{Conclusion and Discussion}
By using Monte Carlo simulation, we demonstrate that half-cylindrical colloidal rods with short-range attraction can self-assemble into two isostatic self-dual crytals with p31m (twisted Kagome) and pmg symmetries. To the best of our knowledge, this special pmg isostatic structure has not been reported before, which exhibits unconventional critical-tilted Dirac cones and rare finite frequency floppy modes. Since half-cylindrical  geometry can be projected by 2D semi-disk pattern, one can apply current mature techniques based on photolithography or mold  to fabricate half-cylindrical rods~\cite{k2017metalenses,k2016metalenses, hao2020vertically}. This method can also be used to directly fabricate 2D array of end-fixed half-cylindrical pillars, where the central-force bonds between pillars are realized by the { screened} electrostatic repulsion. Similar open crystals are also expected to be self-assembled from bowl-shape~\cite{marechal2010phase} or banana-shape~\cite{f2020shaping} colloidal particles with patchy modifications. Our finding opens up new possibilities in designing 2D metameterials with exotic mechanical or phononic properties~\cite{rev2019t,b2017flexible}. For example, the existence of the critically-tilted Dirac cone in this  self-assembled structure suggests that it is possible  to  fabricate  zero-index phononic or mechanical materials  in a cheaper and faster way~\cite{z2020zero,liberal2017near}.  It also provides opportunities in classical systems to observe some  intriguing phenomena which  commonly exist in extreme conditions, e.g., the black hole horizon~\cite{huang2018black,liu2020fermionic,volovik2016black}.

\begin{acknowledgments} 
\subparagraph{Acknowledgments:}  This work is supported by the National Natural Science
Foundation of China (Nos. 11474155 and 11774147), by Singapore Ministry of Education through the Academic Research Fund MOE2019-T2-2-010 and RG104/17 (S), by Nanyang Technological University Start-Up Grant (NTU-SUG: M4081781.120), by the Advanced Manufacturing and Engineering Young Individual Research Grant (A1784C0018) and by the Science and Engineering Research Council of Agency for Science, Technology and Research Singapore. 
\end{acknowledgments}

\bibliographystyle{nature}
\bibliography{reference}

\end{document}